\documentclass[lefteqn,showpacs,preprint]{revtex4}

\begin{document}
\draft
\title{One Way Light Transmission through metal membranes}
\author{Hai Wang,Yun-Song Zhou$^{\dag}$,Shan-Shan Zhang, He Wang,Guo-Zhong Zhao,Zhi-Peng Li,Pei-Jie Wang,Hong-Wei Jiang}
\date{\today }
\draft
\address{Department of Physics, Capital Normal University. Beijing 100048, China}

\begin{abstract}
We propose a new mechanism for irreversibility of light assisted by surface plasmon polaritons (SPPs). By achieving the different mutual conversion efficiency between the radiation field and the SPPs on the opposite surfaces of metal membrane, the unidirectional transmission is observed in a quite simple asymmetric metallic sub-wavelength structures Ag(15nm)/AAO, in which the difference of the two-way transmittance is large than 50\%.
\end{abstract}


\maketitle

\newpage

SPPs excited by electromagnetic wave were firstly proposed by Ritchie [1]. A milestone was made by T. W. Ebbesen \emph{et al.} in 1998 for the extraordinary optical transmission through sub-wavelength holes due to SPPs' excitation on the metal surfaces [2]. In the past ten years, research on SPPs has made great achievements [3].We note in particular some of the work. In 2002, a Bovine structure has been proposed by Ebbesen et al [4],in which the light transmission through a sub-wavelength hole can be well controlled by the micro-structures fabricated around the hole. SPPs are excited more efficiently by periodical concentric circles in the incidence plane. On the exit surface, the same structure can beam the light. It is a brilliant experiment to show a controllable behavior of light by the metallic microstructures. However, let's continue their experiment. Assuming only one side of thin metal films retained concentric grooves, the other side is flat. When the concentric circles exist in the exit plane, the beaming effect is still expected, which benefits the far field energy. While there are no incident surface concentric circles, less SPPs are stimulated, which detriments the intensity of far field. Turn this single side Bovine over, let the concentric circles in incidence plane, more SPPs are excited efficiently, but no beaming effect on another side. Obviously, SPPs excitation and beaming cannot be balanced in single side Bovine. Thus what the influence occurs for the far field intensity? If it does, the unidirectional transmittance comes to a reality. More recently, G. Zheng \emph{et al.} showed that a single side Bovine can be a point filter to achieve angular dependent transmittance [5,6]. However, what kind of angular distribution expected if they turn the Bovine over? If not, again, a unidirectional transmittance structure may be created. All those pioneers' researches [1-20] inspire us to challenge the unidirectional transmission of light based on SPPs.

Let's first consider a metallic sub-wavelength structure. For the time being not consider the details of its structure, it can be either the hole array (or single hole) or slit array (or single slit), but the two faces is asymmetrical. As Fig.1(a) shows, from the left to the right, SPPs excite on the left side with a conversion efficiency $\xi_{R-S}$. Part of them transmits to the right side and the transmittance of SPPs is $t$. On the right side, SPPs convert to the radiation field with an efficiency $\xi_{S-R}$. Thus, the intensity relationship between the source and the detected signal can be simply expressed as $I=I_{0}\xi_{R-S}t\xi_{S-R}$, here, the subscript R-S (S-R) means the transform from the radiation field to SPPs (or vice verse).Based on the same physical picture, in the case of left-to-right as shown in Fig.1(b), we have $I^{'}=I_{0}\xi_{R-S}^{'}t\xi_{S-R}^{'}$, in which $\xi_{R-S}^{'}$($\xi_{S-R}^{'}$)and $t^{'}$ are the transform efficiency and the SPPs' transmittance. The question now is£º$I$ always equals to $I^{'}$? It is true for conventional optics, in which the optical reversibility dominates. Thus the energy is same for two-way transmission. But such judgments are based on the principles of geometrical optics and do not involve the conversion of energy. However, in sub-wavelength optics, the new physics is the  between light and SPPs on the surface of metal. Since the asymmetry of the structure, there is no reason to believe that the conversion efficiency of two surfaces is same. A general case should be:$\xi_{R-S}\xi_{S-R}\not\equiv\xi_{R-S}^{'}\xi_{S-R}^{'}$, thus $I \not\equiv I^{'}$. Even $t=t^{'}$,the above conclusion still holds. Thus a sub-wavelength asymmetry metallic structure is likely to cause a one way transmission. To confirm the above analysis, we prepared a Ag/AAO asymmetric membrane and observed the unidirectional transmission of light.

Fig.2 shows the THz transmittance spectra [21] of Si wafer and Ag(15nm)/Si wafer. In Fig.2, $T_{0}$ represents THz transmittance spectra of Si wafer;$T_{1}$ is for THz transmittance spectra of Ag/Si; $T_{2}$ is for Si/Ag. Here $T_{1}$ and $T_{2}$ are measured the same sample from the opposite sides. As shown in Fig.2, the THz transmittance ratio of Si wafer ($T_{0}$)is no lower than $60\%$,however,the ratio declines largely while 15nm Ag is covered on Si wafer surface ($T_{1}$ and $T_{2}$)since THz wave could not transmit through metal. The inserted one shows the transmittance of Ag/Si from both sides in detail, in which their tiny differences are submerged by the noise. The transmit sequence has no effect on spectra, which is obviously consistent with the traditional optics. For our samples, the roughness of Si wafer surface and Ag thin film surface is quite small (typically < 0.2nm for Ag surface). Thus SPPs is hardly excited at Ag surface and could not be expected to participate with the light transmission.

However, if SPPs play a role in light transmission through a metallic structure, what will be happen for the far field measurement from both sides? Encouraged by those pioneers' works [1-20], a simple structure metal/AAO is fabricated. To emphasize the comparability, a 15nm thick Ag thin layer is deposited on AAO template.

The morphology of AAO template is shown in Fig.3(a), in which the plum piles like microstructure can be clearly seen [22]. However, those details have no effect on THz wave transmission from the two opposite directions. Transmittance of AAO template is quite high as shown in Fig.3(b), which is due to two aspects: 1. the cross-section view measurement by SEM shows that the thickness of AAO template is about $17 \mu m$, which is much thinner than Silicon wafer ($750 \mu m$); 2. a hole array along light transmission direction decreases the light scattering and the absorption of medium. It is noted that a peak appears around 1.7THz which may be explained by the optic interference, however, it is not the main topic discussed in the present paper.

Then, 15nm thick Ag layer is deposited onto AAO template. A significant roughness of Ag surface is observed by SEM, as shown in Fig.4(a). The top surface and the bottom surface of Ag thin film are different. Those micro- or nano- Ag structures offer the "hot spots" for the surface SPPs excitation since Ag is still the good metal in THz range [23]. The transmittance spectra are measured from both sides of Ag/AAO, as shown in Fig.4 (b). The curve $T_{1}$ (or $T_{2}$) corresponds to the case, in which the Ag (or AAO) surface is the incident plane. $T_{1}$ and $T_{2}$ vary in the ranges of $0.8\%¡«1.4\%$, and $0.4\%¡«0.8\%$, respectively£¬in THz region. The discrepancy between $T_{1}$ and $T_{2}$ is significant. In a certain frequency region, the ratio of reaches to about $50\%$.Compared with previous measurements on Ag/Si structure, here, the noise level is much lower than the signal. Furthermore, when two electronic diodes reverse series, any direction of the electric flux will dramatically reduce. Similarly, if the two Ag/AAO structure reverse series, form Ag/AAO/Ag structure, the photon transmission becomes as $T_{12}=T_{21}=T_{1}T_{2}$, here the subscripts 12 and 21 represents the incidence from the opposite sides. The results are also shown in Fig.4(b), which is consistent with our predictions. (Since it is hard to hold a perfect similarity between the top and the bottom surface, tiny discrepancies exist in details).

All above observations and theoretical discussions lead us to believe that one way light travel has been observed in our experiments, which is correct if the discrepancy of conversion efficiency between SPPs and light are different from both sides. Negative aspects for our sample are that transmittance ratio is low and discrepancy of transmittance is still small. However, many pioneers' experiments and theoretical calculations [5-21] have shown that, a high transmittance can be expected by optimize the SPPs device structures.

The electromagnetic energy conversion along metal surface is the subject of the relationship between light radiation field and SPPs. Controllable conversion efficiency between SPPs and far field is one of the important fundamental issues for its applications. Here we show that, even in a simple and rough device, an intrinsic unidirectional light propagation can be achieved. It is no doubt that a much better device with a high unidirectional transmittance can be invented in future by profoundly understanding of the conversion efficiency between SPPs and far field in experiments and theories. The principles of one way transmission properties of sub-wavelength metallic structures in THz band can be extended to the other optical bands. Moreover, based on the unidirectional transmission, a light diode may be created, which will effectively prompt the development of logical optical circuit.

\vskip8pt \textbf{Acknowledgements}

\vskip5pt This work was supported by the National Natural Science Foundation of China (Grant Nos.10904097 and 10874124), as well as PHR(IHLB) from Beijing..

%\newpage
\[
\]
\centerline{\bf Experimental Section}

\begin{description}
\item {AAO template fabrication}: High-purity ($99.999\%$) aluminum foil of $0.2mm$ thickness was used. Before anodization, the aluminum foil was degreased with acetone first, then washed in deionized water. After this cleaning process, the foils were annealed in vacuum at $550¡æ$ for $5$ hours¡£Then the aluminum was electro-polished in a mixture of perchloric acid and ethanol (1:4 in volume) under DC voltage of $18 V$ for $2 min$ at $0¡æ$. Aluminum foils were anodized in a $0.3 M$ oxalic acid solution at 0¡æ under a constant voltage of $45 V$ with $2 hours$. The oxide layer formed in the first step was removed by chromic acid at $60¡æ$. The second anodization time is two hour by using the same electrolyte used in the first anodization. The left aluminum behind AAO template was removed by the mixture solution of perchloric acid ($600ml$ deionized water and $90ml$ perchloric acid) and saturated solution of copper chloride.

\item {Sputtering Condition}: Silver thin film were grown in JGP600 high vacuum system with a base pressure of $1.5\times10^{-5}$ Pa by dc magnetron sputtering method. The deposition rate is 0.1nm/s and the working pressure was $0.5Pa$. The sample holder was well cooled by cycle cooling water.

\item {Morphology Characterization}: The morphology of AAO and Ag/AAO were observed by Hitachi S-4800 field emission scanning electron microscopy.

\item {Terahertz spectra measurement}: The terahertz spectra of all samples are measured by the standard terahertz TDS system. A diode pumped mode-locked Ti: sapphire laser (Mai Tai, Spectra Physics) with the repetition rate of $82MHz$ provides the femtosecond pulses with the duration of $100fs$ and the center wavelength of $800nm$, and the average output power of $1.08W$. A femtosecond laser pulse is divided into two beams. One is the pump beam to generate terahertz radiation. Another is the probe beam to monitor the temporal terahertz field. Four parabolic mirrors are used to collimate and focus terahertz beam through free space onto the sensor crystal. A balanced photodiode differential detector measures the probe beam after the sensor crystal and Wollaston prism. The signal is collected and amplified by a lock-in amplifier and acquired by the computer for the display and data processing.

\item {n-type Si(111) wafer}:  Resistivity: $\rho > 3000 \Omega\cdot cm$  ; Thickness: $700\mu m$.

\end{description}

%\newpage

\centerline{\bf References}
%begin{references}
%\bibitem{}
%\bibitem{}
%\vspace{-1.6cm}
%\begin{description}
%\end{description}

%\end{references}

%\newpage
\[
\]
\centerline{\bf Figure captions}

\begin{description}
\item {Fig. 1} The sketch of the light transmission through a sub-wavelength metal structure. (a) from the right side to the left; (b) from the left side to the right.

\item {Fig. 2} THz transmittance spectra of Si wafer and Ag(15nm)/Si.   represents THz transmittance spectra of Si wafer ($700\mu m$);  $T_{1}$ is for THz transmittance spectra of Ag(15nm)/Si($700\mu m$);  $T_{2}$ is for Si($700\mu m$)/Ag(15nm). Here $T_{1}$ and $T_{2}$ are measured the same sample from the opposite sides. The inserted one is the enlarged plot of Ag/Si THz transmittance from both sides.

\item {Fig. 3} SEM image of AAO template and its THz transmittance spectra.

\item {Fig. 4} The surface morphology of Ag/AAO; (b)  $T_{1}$ (black curve) is for THz transmittance spectra of Ag(15nm)/AAO($17\mu m$),;  $T_{2}$ (red curve) is for AAO ($17\mu m$)/Ag(15nm),. $T_{12}$(=$T_{21}$)  (blue one) represents Ag(15nm)/AAO/Ag(15nm). $T_{1}$ and $T_{2}$ are the THz transmittance spectra measured for the same sample from opposite sides.

\end{description}

%\end{multicols}

\end{document}